\documentclass[runningheads]{llncs}
\usepackage[T1]{fontenc}
\usepackage{graphicx}
\usepackage{dsfont}
\usepackage[table,xcdraw]{xcolor}
\definecolor{mylinkcolor}{RGB}{73,49,49}
\definecolor{myrefcolor}{RGB}{124, 75, 0}
\usepackage[colorlinks = true,
   linkcolor = blue,
   urlcolor  = black,
   citecolor = blue,
   anchorcolor = blue]{hyperref}
\usepackage{amsmath}
\usepackage{amssymb}
\usepackage{multirow}
\usepackage{booktabs}
\usepackage{footnote}
\usepackage{tablefootnote}
\usepackage{soul}
\usepackage{algorithm}
\usepackage{algorithmic}
\usepackage{wrapfig}
\usepackage{subcaption}
\usepackage{array}
\usepackage{pifont}
\usepackage{cite}
\usepackage{marvosym}
\newcolumntype{I}{!{\vrule width 1.2pt}}
\makeatletter
\newcommand{\printfnsymbol}[1]{%
  \textsuperscript{\@fnsymbol{#1}}%
}
\usepackage{lipsum}

\newcommand\blfootnote[1]{%
  \begingroup
  \renewcommand\thefootnote{}\footnote{#1}%
  \addtocounter{footnote}{-1}%
  \endgroup
}
\makeatother
\usepackage{color}

\begin{document}
\title{A Foundation Model for Brain Lesion Segmentation with Mixture of Modality Experts}

\author{Xinru Zhang$^{\mathrm{1},\mathrm{2},\textcolor{blue}{*}}$, Ni Ou$^{\mathrm{3}}$, Berke Doga Basaran$^{\mathrm{4}}$, Marco Visentin$^{\mathrm{4}}$, \\ Mengyun Qiao$^{\mathrm{2}}$, Renyang Gu$^{\mathrm{5}}$, Cheng Ouyang$^{\mathrm{4}}$, \\Yaou Liu$^{\mathrm{6}}$, Paul M. Matthews$^{\mathrm{2,7}}$, Chuyang Ye$^{\mathrm{1},}$\textsuperscript{\href{mailto:chuyang.ye@bit.edu.cn}{\textcolor{blue}{\Letter}}}, Wenjia Bai$^{\mathrm{2,4,}}$\textsuperscript{\href{mailto:w.bai@imperial.ac.uk}{\textcolor{blue}{\Letter}}}}
\institute{$^{1}$School of Integrated Circuits and Electronics, Beijing Institute of Technology, Beijing, China\\ 
$^{2}$Department of Brain Sciences, Imperial College London, London, UK\\
$^{3}$School of Automation, Beijing Institute of Technology, Beijing, China\\
$^{4}$Department of Computing, Imperial College London, London, UK\\
$^{5}$Department of Bioengineering, Imperial College London, London, UK \\
$^{6}$Beijing Tiantan Hospital, Capital Medical University, Beijing, China \\
$^{7}$UK Dementia Research Institute, Imperial College London, London, UK 
}

\titlerunning{MoME: Mixture of Modality Experts}
%
%
\authorrunning{Xinru Zhang et al.}
%
%
\maketitle 
\begin{abstract}
Brain lesion segmentation plays an essential role in neurological research and diagnosis. 
As brain lesions can be caused by various pathological alterations, different types of brain lesions tend to manifest with different characteristics on different imaging modalities.
Due to this complexity, brain lesion segmentation methods are often developed in a task-specific manner. A specific segmentation model is developed for a particular lesion type and imaging modality.
However, the use of task-specific models requires predetermination of the lesion type and imaging modality, which complicates their deployment in real-world scenarios.
In this work, we propose a universal foundation model for 3D brain lesion segmentation, which can automatically segment different types of brain lesions for input data of various imaging modalities. 
We formulate a novel \textit{Mixture of Modality Experts}~(MoME) framework with multiple expert networks attending to different imaging modalities. 
A hierarchical gating network combines the expert predictions and fosters expertise collaboration. Furthermore, we introduce a curriculum learning strategy during training to avoid the degeneration of each expert network and preserve their specialisation. 
We evaluated the proposed method on nine brain lesion datasets, encompassing five imaging modalities and eight lesion types.
The results show that our model outperforms state-of-the-art universal models and provides promising generalisation to unseen datasets.

\keywords{Foundation Model \and Brain Lesion Segmentation \and Mixture of Experts}
\end{abstract}
\blfootnote{\textcolor{blue}{*} Work conducted as a visiting PhD student at Imperial College London, under the joint supervision of \textsuperscript{\textcolor{blue}{\Letter}}corresponding authors.}
\section{Introduction}

\begin{figure}[!t]
    \centering
    \includegraphics[width=\textwidth]{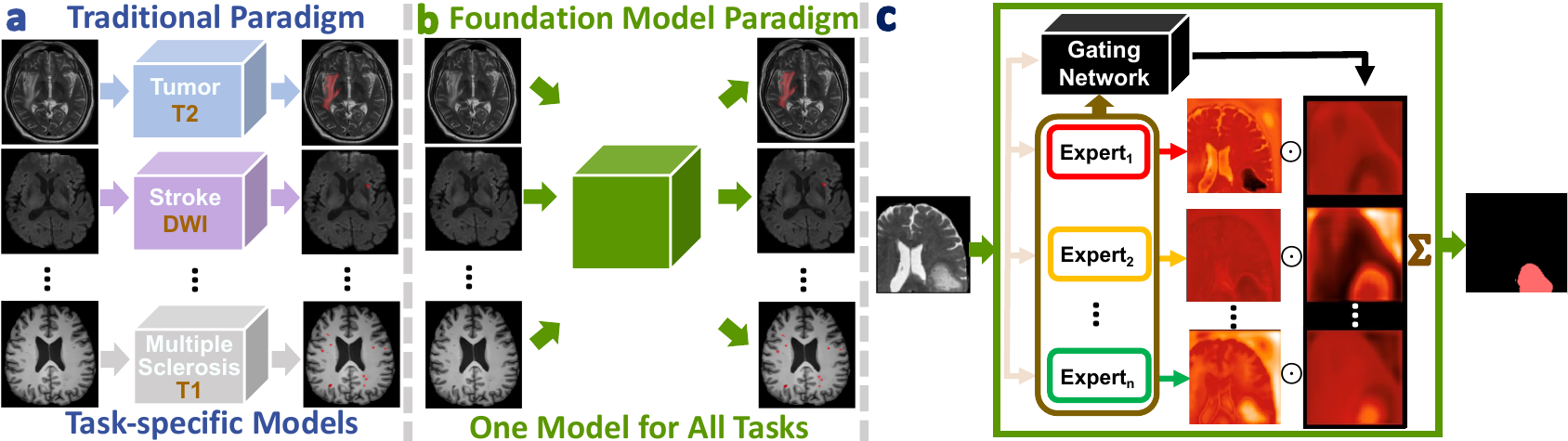}
    \caption{Different paradigms for brain lesion segmentation: a) the traditional paradigm that trains multiple task-specific models; b) the foundation model paradigm that trains a single universal model for multiple tasks; c) the proposed \textit{mixture of modality experts} 
    ($\text{MoME}$) framework for constructing the foundation model.}
    \label{fig:foundation model}
\end{figure}

Brain lesion segmentation plays an essential role in neurological research and diagnosis by offering a quantitative description of pathologies~\cite{kamnitsas2017efficient,zhang2023carvemix}, and deep learning has greatly advanced the progress of brain lesion segmentation in recent years~\cite{wood2022deep,czolbe2023neuralizer,basaran2023lesionmix}. 
Since brain lesions tend to manifest with different characteristics due to various patient-dependent pathological alterations, the segmentation of different types of brain lesions requires the use of different \textit{magnetic resonance imaging} (MRI) modalities~\cite{shah2013discriminating,wu2006characterizing,schmidt2007high}, which increases the segmentation complexity.
In the traditional paradigm (Fig.~\ref{fig:foundation model}a), separate task-specific models are trained to segment specific types of brain lesions on the corresponding MRI modalities that are sensitive to the lesions~\cite{zhang2023carvemix}.
This paradigm necessitates the development of numerous models for clinical analysis tasks~\cite{ulrich2023multitalent,gao2023training}. Clinicians are compelled to manually choose the optimal model for each patient-specific MRI scan based on predetermined brain lesion type criteria. This process hinders practical deployment of the \textit{artificial intelligence}~(AI) models.
Contrary to this traditional paradigm, we propose a task-agnostic universal foundational model (Fig.~\ref{fig:foundation model}b), which is capable of automatically segmenting various types of brain lesions based on inputs of different MRI modalities.

Foundation models have gained popularity in medical image segmentation, thanks to the success of the \textit{Segment Anything Model}~(SAM)~\cite{kirillov2023segment} and its variants~\cite{wang2023sam,gong20233dsam,ma2024segment}, as well as other innovative foundation models~\cite{zhou2023foundation,ulrich2023multitalent,gao2023training,butoi2023universeg} specifically developed for medical imaging. 
Earlier works aim to develop a universal model across all modalities, organs, and tasks. 
However, due to the complexity of medical imaging data and tasks, the segmentation accuracy of such a universal model is often limited~\cite{zhang2023challenges}. 
Thus, foundation models focusing on narrower scopes are explored, which are designed for a particular modality or a particular organ/task~\cite{zhang2023challenges}. 
For example, several models are proposed specifically for \textit{computed tomography}~(CT) image segmentation~\cite{wasserthal2023totalsegmentator,liu2023clip,ulrich2023multitalent}, showing promising results for anatomical structure and large tumour segmentation. 

Despite these progresses, most existing foundation models may not be directly applicable to brain lesion segmentation.
Most noticeably, universal brain lesion segmentation requires the capability of handling the challenge of input modality diversity, which is not addressed by many of the existing models.
SAM-inspired methods, such as SAM-Med3D~\cite{wang2023sam} and MedSAM~\cite{ma2024segment}, require human-given prompts, which can be impractical for brain lesion segmentation as the lesions can be scattered across the brain.
Neuralizer~\cite{czolbe2023neuralizer} demonstrates the potential of a foundation model for brain imaging that generalises to new tasks by learning transferable features from a context set to the input image. It is evaluated on 2D coronal slices, which might limit its use on 3D brain lesion segmentation task. 
Hermes~\cite{gao2023training} is proposed with the awareness of the modality diversity, where modality information is embedded as a prior in model training.
However, it does not fully explore the relationship between different modalities. The development of a universal foundational model for brain lesion segmentation with diverse input modalities is still an open problem.

In this study, we propose a \textit{Mixture of Modality Experts}~(MoME) model, a foundation model for brain lesion segmentation capable of handling diverse imaging modalities and accommodating various types of brain lesions.
As shown in Fig.~\ref{fig:foundation model}c, inspired by \textit{Mixture of Experts}~(MoE)~\cite{ou2022patcher,chen2023adamv,zhu2022uni}, MoME enhances the model capacity by assembling a team of experts, where each expert is specialised to handle a particular imaging modality. 
A hierarchical gating network is developed to combine the expert predictions and foster collaborative expertise exploration. 
Moreover, to avoid the degeneration of each expert network and preserve their specialisation, we introduce a curriculum learning strategy for training the MoME model. 
For evaluation, we conducted comphrehensive experiments on nine brain imaging datasets comprising a total number of 6,585 annotated 3D images, with five MRI modalities and eight lesion types. The results show that MoME outperforms competing methods in multiple aspects and provides improved generalisation to unseen datasets. The codes of our method are available at \url{https://github.com/ZhangxinruBIT/MoME}.

\section{Methods}

\subsection{Method Overview}

\textbf{The architecture of MoME} features two major components, namely the expert networks and the gating network. 
Each expert network is adept in a specific domain of imaging modality, contributing to {expert specialisation}. The gating network functions as a decision-maker and ensembles the expert networks by dynamically adjusting the weights of their outputs with {hierarchical guidance}.
\textbf{Regarding the training of MoME}, an {anti-degeneration strategy} based on curriculum learning is developed, where model training gradually transitions from specialising the role of each expert to determining the entire model through collaboration.
\subsection{Detailed Design}

\subsubsection{Expert Specialisation}
\label{Sec:expert_def}
Suppose there are $N$ imaging modalities $\{\mathcal{M}_i\}_{i=1}^{N}$ of interest, where $\mathcal{M}_i$ represents the $i$-th modality.
To address the problem of modality diversity in brain lesion segmentation, in MoME, $N$ expert networks~$\{{E}_{i}\}_{i=1}^{N}$ are developed, where $E_{i}$ denotes the $i$-th expert. 
Each expert $E_{i}$ is expected to specialise in a specific modality $\mathcal{M}_i$.
Without losing generality, we assume that each expert network follows a multi-resolution encoder-decoder architecture, which is common for medical image segmentation such as in U-Net~\cite{isensee2021nnu}. 
We focus on the five most widely used brain imaging modalities, including T1-weighted, T2-weighted, T1 contrast-enhanced (T1ce), FLAIR, and \textit{diffusion weighted imaging}~(DWI), and thus $N=5$.
Note that the experts here are built for 3D brain image segmentation, distinguishing them from sparse MoE architectures for 2D natural images where a higher number of experts can be afforded and routing is employed for expert selection~\cite{puigcerver2023sparse,zhang2023robust}. Due to the smaller number of experts, we do not explore expert routing in this work.

\subsubsection{Hierarchical Gating Network}
\label{sec:Layer-Wise Gating}
Following the specialisation of expert networks, we propose a gating network \(G\) that facilitates the collaboration and conducts voxel-wise weighted aggregation among the expert outputs. 
Note that different imaging modalities can be correlated and provide complementary information. For example, FLAIR appears similar to T2 but with fluid being attenuated. They may both contribute to lesion segmentation with different  features,  and the outputs of an expert that is not specialized in the input modality may still provide complementary useful information.
To enhance collaboration, the gating network is hierarchical  and leverages multi-resolution information jointly.

Mathematically, given an input image $\mathbf{X}$, each expert ${E}_{i}$ generates outputs at multiple resolution levels on the decoder side, denoted by ${E}_{i}(\mathbf{X}) = \{{\mathbf{e}_{i}^{l}}\}_{l=1}^{L}$, where $\mathbf{e}_{i}^{l}$ is the output of ${E}_{i}$ at the $l$-th resolution level and $L$ indicates the number of resolution levels. 
The gating network ${G}$ generates a set of hierarchical voxel-wise weight maps at different resolution levels for each expert, denoted as $\{{\mathbf{g}_{i}^{l}}\}_{l=1}^{L}$.
Combining the expert networks and the gating network, the final output $\mathbf{O}$ is formulated as
\begin{equation}
\mathbf{O} = \left\{\mathbf{o}^l\right\}_{l=1}^{L} \quad\text{with}\quad \mathbf{o}^l= \sum_{i=1}^{N} \mathbf{e}_i^l \odot \mathbf{g}_i^l,
\label{eq:}
\end{equation}
where $\mathbf{o}^l$ is the aggregated output of all experts at the $l$-th level and $\odot$ represents voxel-wise multiplication. 
The segmentation at the $l$-th level is obtained by applying the softmax operator to $\mathbf{o}^l$.
During training, deep supervision~\cite{lee2015deeply,isensee2021nnu} is employed to guide the segmentation at multiple resolution levels. At the inference stage, only the highest resolution level determines the final segmentation result~\cite{lee2015deeply,isensee2021nnu}.

\subsubsection{Curriculum Learning to Alleviate Expert Degeneration}
\label{sec.curriculum_learning}

Although it is feasible to simply jointly train the experts and gating network, this may lead to expert degeneration, where each expert no longer focuses on a specific modality and their activation becomes imbalanced~\cite{chi2022representation,chen2023mod}, reducing the capability of the whole model to handle diverse modalities.
To address this issue, we propose a curriculum learning strategy for model training, which gradually transitions from specialising the experts to determining the whole model, including gating-based expert collaboration and expert refinement.

Specifically, the curriculum learning is achieved with a dynamic loss $\mathcal{L}_{\mathrm{cl}}$ that comprises the specialisation loss $\mathcal{L}_{i}$ and the collaboration loss $\mathcal{L}_{\text{MoME}}$:
\begin{equation}
\label{eq:L_cl}
\mathcal{L}_{\mathrm{cl}} = f_{\text{epoch}} \cdot \sum_{i=1}^{N} \mathds{1}_{\mathbf{X}\in\mathcal{M}_i} \cdot  \mathcal{L}_{i}  + (1-f_{\text{epoch}}) \cdot \mathcal{L}_{\text{MoME}}.
\end{equation}
Here, $\mathds{1}_{\mathbf{X}\in\mathcal{M}_i}$ is an indicator function that is one when the input image $\mathbf{X}$ belongs to the modality $\mathcal{M}_{i}$ and zero otherwise, and $f_{\text{epoch}}$ is a dynamic weight that changes with the epoch iteration.
$f_{\text{epoch}}$ is designed as $f_{\text{epoch}} = (1-\text{epoch}_{\text{current}}/\text{epoch}_{\text{total}})^{2}$, where $\text{epoch}_{\text{current}}$ and $\text{epoch}_{\text{total}}$ are the current epoch and total number of epochs, respectively.
In this way, during the early training stage, $f_{\text{epoch}}$ is greater, which emphasises the loss $\mathcal{L}_{i}$ ($i\in\{1,\ldots,N\}$) for each expert $E_{i}$ that matches the modality of $\mathbf{X}$.
During the later training stage, $f_{\text{epoch}}$ decreases and $\mathcal{L}_{\text{MoME}}$ becomes gradually more emphasised.

$\mathcal{L}_i$ is designed with deep supervision at each resolution level as
\begin{equation}
\label{eq:Li}
\mathcal{L}_i = \sum_{l=1}^{L} k^l \cdot \mathcal{L}(\sigma(\mathbf{e}_i^l),d^l(\mathbf{Y})),
\end{equation}
where $k_{l}$ is a weight for the $l$-th level, $\sigma(\cdot)$ is the softmax activation, \(d^l(\cdot)\) is a down-sampling operator that matches the resolution of the annotation \(\mathbf{Y}\) with that of the \(l\)-th level, and $\mathcal{L}(\cdot,\cdot)$ is the combination of the soft Dice loss and cross-entropy loss~\cite{isensee2021nnu}.
$\mathcal{L}_{\text{MoME}}$ is formulated with the collaboration between experts as
\begin{equation}
 \mathcal{L}_{\text{MoME}} = \sum_{l=1}^{L} k^l \cdot \mathcal{L}(\sigma(\mathbf{o}^l),d^l(\mathbf{Y})).
\label{eq:L}
\end{equation}


\subsection{Implementation Details}

We implement each expert network and the gating network with the nnU-Net architecture~\cite{isensee2021nnu}, which is a strong segmentation backbone as shown in~\cite{isensee2021nnu,ulrich2023multitalent}.  
The input to the gating network is a combination of raw images and shallow features extracted from the first encoder layers of the experts.
As each expert network is expected to specialise in one imaging modality, the expert networks are first trained separately to acquire the modality knowledge. 
Then, the gating network is trained jointly with the fine-tuning of the expert networks based on the dynamic curriculum learning loss Eq.~\ref{eq:L_cl}.
All the hyperparameters~(including $\{k_l\}_{l=1}^{L}$) follow the default setting in nnU-Net~\cite{isensee2021nnu}, except that 1) a low learning rate of $1e^{-3}$ is set for the dynamic curriculum learning loss~\cite{rajbhandari2022deepspeed} and 2) the number of training epochs is set to 800, a sufficient number for convergence.
\section{Experiments}
\subsection{Dataset and Task Description}
We curated six publicly available brain lesion segmentation datasets (\href{https://portal.fli-iam.irisa.fr/msseg-challenge/}{MSSEG}~\cite{MSSEGcommowick2018objective}, \href{https://www.oasis-brains.org/}{OASIS}~\cite{OASISmarcus2010open}, {\href{https://www.isles-challenge.org/}{ISLES2022}~\cite{ISLEShernandez2022isles}}, \href{http://www.brainTumoursegmentation.org/}{BraTS2021}~\cite{BraTSbaid2021rsna}, \href{https://wmh.isi.uu.nl/\#_Toc122355653}{WMH2017}~\cite{WMH2017kuijf2019standardized}, and {\href{https://fcon_1000.projects.nitrc.org/indi/retro/atlas.html}{ATLAS2.0}~\cite{ATLASliew2018large}})
and three in-house datasets (Tumour1, Tumour2, and WMHsMix), amounting to a total number of 6,585 3D images with brain lesion annotations. The dataset detail is reported in Supple. Table~{\bfseries\textcolor{orange}{\uppercase\expandafter{\romannumeral2}}}.
All images were registered to the MNI152 template using {\href{https://stnava.github.io/ANTs/}{ANTs}}~\cite{avants2011reproducible}. Then, brain masks were applied to the images for skull removal. Cropping was performed so that the size of all images was standardised to $160\times196\times160$ with a voxel size of $1\times1\times1~\mathrm{mm}^3$. Each of the six public datasets was split into training and test sets. The MoME model was trained on the combined training sets from all six public datasets. It was then evaluated on the test sets from the public datasets (\textbf{seen datasets}) and the three in-house datasets (\textbf{unseen datasets}), which were never used for training and allowed the assessment of the generalisation of MoME. 
 
\label{sec:experiment settings}

\subsection{Competing Task-specific and Foundation Models}

MoME was compared with the following task-specific and foundation models.

\textbf{Task-specific nnU-Nets}: The nine datasets cover a variety of brain lesions and modalities. We regard the segmentation of each modality for each dataset as a distinct task~\cite{butoi2023universeg}, resulting in a total number of 17 tasks~(14 tasks for seen datasets and 3 tasks for unseen datasets; task IDs listed in Supple. Table~{\bfseries\textcolor{orange}{\uppercase\expandafter{\romannumeral1}}}). Thus, 17 supervised task-specific nnU-Net models were trained.

\textbf{nnU-Net}: A single nnU-Net~\cite{isensee2021nnu} was trained with the combined training sets from all public datasets. It can be viewed as a baseline foundation model.

\textbf{Multi-Talent}~\cite{ulrich2023multitalent}: Multi-Talent demonstrates superior performance in handling conflicting class definitions for multi-dataset training and shows promising lesion segmentation results. We trained it with the combined training sets.

\textbf{Hermes}~\cite{gao2023training}: Hermes is a foundation model that injects context prior knowledge (e.g. anatomical region and imaging modality) into the segmentation network to address the challenges of data and modality diversities. We adapted Hermes for our task and incorporated the imaging modality information as prior knowledge. Hermes was also trained with the combined training sets.

\textbf{SAM-Med3D}~\cite{wang2023sam}: This is a SAM-based~\cite{kirillov2023segment} model trained on a large-scale 3D medical image dataset. We fine-tuned {\href{https://drive.google.com/file/d/1MuqYRQKIZb4YPtEraK8zTKKpp-dUQIR9/view}{SAM-Med3D} with the combined training sets and provided a 10-point prompt~\cite{wang2023sam} during the inference.

All models were trained for 800 epochs on Nvidia A100 GPUs. 
We used the default hyperparameters for the transformer-based SAM-Med3D. All the other models were based on nnU-Net and used its default hyperparameters. 

\subsection{Results}

\subsubsection{Performance on Seen Datasets}
Table~\ref{tab:seen} compares the average Dice score of MoME with the competing models and reports the overall GPU memory usage for training each model.
To mitigate the bias caused by the varying numbers of images in different test sets, we report the Dice score with different levels of averaging: across the 14 seen tasks~(task-level), across 6 seen datasets~(dataset-level), and across 1,230 test images~(image-level). 
MoME outperforms the other foundation models, for example, achieving 2\% to 4\% higher Dice scores than nnU-Net, Multi-Talent, and Hermes. 
It outperforms SAM-Med3D by a larger margin of 7\% to 15\%. This is possibly because SAM-Med3D requires point prompts to indicate a foreground region, which may not be suitable for brain lesion segmentation, where lesions can be numerous, small, and scattered across the brain, such as \textit{white matter hyperintensity}~(WMH) lesions~\cite{WMH2017kuijf2019standardized,basaran2023lesionmix}. The performance of MoME is comparable to task-specific nnU-Nets. However, the latter requires the training of 14 separate models, and its total amount of GPU memory consumption is about three times that of MoME. 

\begin{table}[!t]
\centering
\caption{Average Dice score of MoME and the competing methods on seen datasets. The best results among the foundation models are highlighted in bold purple. The GPU memory usage for model training is also displayed.}
\resizebox{0.88\textwidth}{!}{
\begin{tabular}{clcccccc}
\toprule
\multicolumn{2}{c}{{ }}  & \multicolumn{5}{c}{{ Foundation Model}}& { }  \\ \cline{3-7}
\multicolumn{2}{c}{\multirow{-2}{*}{{ Dice}}}& { nnU-Net } & { SAM-Med3D } & { Multi-Talent } & { Hermes } & { MoME }& \multirow{-2}{*}{{ \begin{tabular}[c]{@{}c@{}}Task-specific\\ nnU-Nets\end{tabular}}} \\ \midrule
\multicolumn{2}{c}{{ Image-level}}  & { 0.8007}& { 0.7322}     & { 0.7948} & { 0.8017} & {\color[HTML]{6200C9} \textbf{0.8204}} & { 0.8202}  \\
\multicolumn{2}{c}{{ Task-level}}& { 0.6654}& { 0.5249}     & { 0.6825} & { 0.6720} & {\color[HTML]{6200C9} \textbf{0.7026}} & { 0.7099}  \\
\multicolumn{2}{c}{{ Dataset-level}}& { 0.6561}& { 0.5083}     & { 0.6681} & { 0.6603} & {\color[HTML]{6200C9} \textbf{0.6938}} & { 0.6950}  \\
\hline
\multicolumn{2}{c}{\begin{tabular}[c]{@{}c@{}}GPU memory \\ usage (GB)\end{tabular}} & { 7.5} & { 32.5}     & { 11.9} & { 7.5}  & { 38.2} & { 104.4} \\ \bottomrule
\end{tabular}
}
\label{tab:seen}
\end{table}
\begin{figure}[!t]
    \centering
    \includegraphics[width=\textwidth]{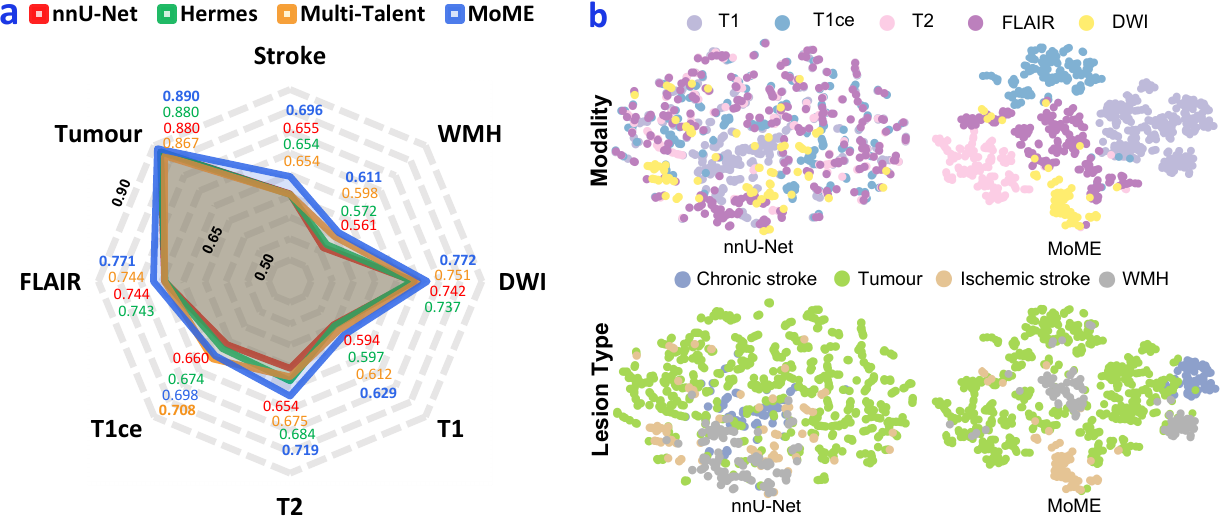}
    \caption{More detailed analysis of the MoME result on seen datasets. a) A radar chart that compares the average Dice score of foundation models from the perspectives of different modalities and lesion types. b) t-SNE plots of latent spaces for nnU-Net and  $\text{MoME}$, where each dot represents a brain image.}
    \label{fig:Radar_TSNE}
\end{figure}
A more detailed analysis of the results is shown in Fig.~\ref{fig:Radar_TSNE}. Fig.~\ref{fig:Radar_TSNE}a presents a comprehensive comparison between the foundation models from 8 more perspectives, reporting the average Dice score for different modalities and different lesion types. 
For 7 out of the 8 perspectives, MoME~(blue curve) outperforms the other foundation models~(SAM-Med3D is not shown due to relatively lower performance but reported in Supple. Table \textcolor{orange}{\textbf{I}}).
Fig.~\ref{fig:Radar_TSNE}b analyses the latent features obtained at the bottleneck of the gating network in $\text{MoME}$ and the single nnU-Net using t-SNE plots~\cite{van2008visualizing}. 
It shows that MoME achieves a more discriminative clustering for images of different modalities and lesion types compared to the nnU-Net, which may be relevant to its improved performance.

\subsubsection{Generalisation to Unseen Datasets}

Table~\ref{tab:dataset level} reports the segmentation performance on the three unseen in-house datasets for evaluating model generalisability. Multi-Talent is not included as it cannot be directly applied to unseen datasets. 
Again, MoME outperforms other competing foundation models. On two unseen datasets (Tumour1 and Tumour2), it even exceeds the task-specific nnU-Nets, which can be considered the upper-bound benchmark as they had access to the training data from the unseen datasets. On the other unseen dataset (WMHsMix), MoME still performs relatively well and achieves an average Dice score of 0.7015, where the dataset contains diverse WMH lesions of five diseases~(see Supple. Table~{\bfseries\textcolor{orange}{\uppercase\expandafter{\romannumeral2}}}). 
\begin{table}[!t]
\begin{minipage}{.51\linewidth}
\caption{Average Dice scores on unseen datasets.
The top scores are in bold purple.}
\label{tab:dataset level}
\resizebox{\textwidth}{!}{
\begin{tabular}{cccccc}
\toprule
     & \multicolumn{4}{c}{Foundation Model}&     \\ \cline{2-5}
     & \multicolumn{1}{l}{}& \multicolumn{1}{l}{}    & \multicolumn{1}{l}{} & \multicolumn{1}{l}{}  &     \\
\multirow{-3}{*}{Datasets}     & \multicolumn{1}{l}{\multirow{-2}{*}{nnU-Net}} & \multicolumn{1}{l}{\multirow{-2}{*}{\begin{tabular}[c]{@{}l@{}}SAM-\\ Med3D\end{tabular}}} & \multicolumn{1}{l}{\multirow{-2}{*}{Hermes}} & \multicolumn{1}{l}{\multirow{-2}{*}{ MoME}} & \multirow{-3}{*}{\begin{tabular}[c]{@{}c@{}}Task-specific\\ nnU-Nets\end{tabular}} \\ \midrule
{ Tumour1} & { 0.8358}  & { 0.8339} & { 0.8446} & {\color[HTML]{6200C9} \textbf{0.8545}}     & { 0.8518} \\
{ Tumour2} & { 0.7959}  & { 0.8146} & { 0.7968} & {\color[HTML]{6200C9} \textbf{0.8293}}     & { 0.8157} \\
{ WMHsMix} & { 0.7032}  & { 0.5479} & { 0.6968} & { 0.7015}     & {\color[HTML]{6200C9} \textbf{0.7508}}   \\ \bottomrule
\end{tabular}
}
\end{minipage}%
\noindent
\hfill
\begin{minipage}{.47\linewidth}

\caption{Ablation studies of MoME. The best scores are in bold.}
\label{tab:ablation}
\resizebox{\textwidth}{!}{
\begin{tabular}{ccllcccc}
\toprule
\multicolumn{4}{c}{{Modality experts}}   & {\color[HTML]{009901} \ding{52}}& {\color[HTML]{009901} \ding{52}}& {\color[HTML]{009901} \ding{52}}& {\color[HTML]{9A0000} \ding{56}}\\
\multicolumn{4}{c}{{Hierarchical gating}}& {\color[HTML]{009901} \ding{52}}& {\color[HTML]{009901} \ding{52}}& {\color[HTML]{9A0000} \ding{56}}& {\color[HTML]{9A0000} \ding{56}}\\
\multicolumn{4}{c}{{Curriculum learning}}  & {\color[HTML]{009901} \ding{52}}& {\color[HTML]{9A0000} \ding{56}}& {\color[HTML]{9A0000} \ding{56}}& {\color[HTML]{9A0000} \ding{56}}\\ \midrule
& \multicolumn{3}{c}{{ Image-level}}   & { \textbf{0.8204}} & { 0.8142} & { 0.8117} & { 0.8027} \\
  & \multicolumn{3}{c}{Task-level}   & { \textbf{0.7026}} & { 0.6969} & { 0.6878} & { 0.6529} \\
\multirow{-3}{*}{\begin{tabular}[c]{@{}c@{}}Seen\\ Datasets\end{tabular}} & \multicolumn{3}{c}{{ Dataset-level}} & { \textbf{0.6938}} & { 0.6852} & { 0.6777} & { 0.6488} \\ \midrule
\multicolumn{4}{c}{Unseen Datasets}  & { \textbf{0.7951}} & { 0.7889} & { 0.7912} & { 0.7814} \\ \bottomrule
\end{tabular}
}

\end{minipage}
\end{table}
\subsubsection{Ablation Studies}

Ablation studies were performed on the seen and unseen datasets to verify the contributions of the designs in MoME with results presented in Table~\ref{tab:ablation}.
For the modality experts, {\color[HTML]{9A0000}\ding{56}}\textbf{/}{\color[HTML]{009901} \ding{52}} 
denotes randomly initialised experts versus specialised experts based on the imaging modality; for the hierarchical gating, {\color[HTML]{9A0000}\ding{56}}\textbf{/}{\color[HTML]{009901} \ding{52}} denotes the basic gating without hierarchical structure versus our hierarchical gating; and for the curriculum learning, {\color[HTML]{9A0000}\ding{56}}\textbf{/}{\color[HTML]{009901} \ding{52}} denotes the application of Eq.~(\ref{eq:L}) versus Eq.~(\ref{eq:L_cl}) during model training.
When the three contributions are sequentially removed, the segmentation performance tends to decrease gradually, which confirms the benefits of these contributions.

\subsubsection{Experts Degeneration}

We further investigated the issue of expert degeneration by comparing the expert output, weighted by gating, without and with the proposed curriculum learning strategy for avoiding degeneration. The detailed results are given in Supple. Fig.~\textcolor{orange}{\textbf{I}}. The figure shows that without the curriculum learning strategy, certain experts can become inactive, losing their modality-specific expertise. The proposed curriculum learning strategy ensures that the experts remain sensitive to modality-specific information.
\section{Conclusion}

We have developed a novel universal foundation model MoME for brain lesion segmentation, where a mixture of modality experts is incorporated to address the challenge of modality diversity. MoME enables collaboration between experts with hierarchical gating and mitigates expert degeneration with a curriculum learning strategy.
Extensive experimental results on nine public and in-house datasets and 17 tasks show that it outperforms competing foundation models across different modalities and lesion types. It is also more efficient than task-specific models, providing a flexible way for model deployment.
\subsubsection{Acknowledgements}
C. Ye is supported by the Beijing Municipal Natural Science Foundation (7242273). 
W. Bai is co-funded by EPSRC DeepGeM Grant (EP/W01842X/1) and NIHR Imperial Biomedical Research Centre (BRC). The views expressed are those of the authors and not necessarily those of the NIHR or the Department of Health and Social Care.
\bibliographystyle{splncs04}
\bibliography{main}
\end{document}


\title{Supplementary Material: A Foundation Model for Brain Lesion Segmentation with Mixture of Modality Experts}
\author{}
\authorrunning{}
%
\institute{}
\titlerunning{Supplementary Material:MoME}
\maketitle   
%
\begin{center} 
\begin{table}[h]
\centering
\caption{Detailed average Dice scores of MoME and competing methods for each task, lesion type, and modality.
The top results, excluding the task-specific nnU-Nets results, are highlighted in purple.}
\resizebox{\textwidth}{!}{
\begin{tabular}{|c|ccc|c|c|c|c|c|c|}
\hline
{ }     & \multicolumn{3}{c|}{{ }}     &    & { } & { }  & { } &          &  \\
{ }     & \multicolumn{3}{c|}{\multirow{-2}{*}{{ Task}}} &    & { } & { }  & { } &          &  \\ \cline{2-4}
{ }     & \multicolumn{1}{c|}{{ }}  & \multicolumn{1}{c|}{{ }}& &    & { } & { }  & { } &          &  \\
\multirow{-4}{*}{{ Settings}}         & \multicolumn{1}{c|}{\multirow{-2}{*}{{ Dataset}}} & \multicolumn{1}{c|}{\multirow{-2}{*}{{ Modality}}} & \multirow{-2}{*}{\begin{tabular}[c]{@{}c@{}}Task\\ ID\end{tabular}}     & \multirow{-4}{*}{\begin{tabular}[c]{@{}c@{}}nnU-Net\end{tabular}} & \multirow{-4}{*}{{SAM-Med3D}} & \multirow{-4}{*}{{ \begin{tabular}[c]{@{}c@{}}Multi-\\Talent\end{tabular}}} & \multirow{-4}{*}{{ Hermes}} & \multirow{-4}{*}{MoME}     & \multirow{-4}{*}{\begin{tabular}[c]{@{}c@{}}Task\\ Specific \\ nnU-Nets\end{tabular}} \\ \hline
    & \multicolumn{1}{c|}{}       & \multicolumn{1}{c|}{T1}      & 1  & 0.8344& 0.7908& 0.8192       & 0.8350& {\color[HTML]{6434FC} \textbf{0.8510}} & 0.8573       \\ \cline{3-10} 
    & \multicolumn{1}{c|}{}       & \multicolumn{1}{c|}{T2}      & 2  & 0.8868& 0.8599& 0.8690       & 0.8893& {\color[HTML]{6434FC} \textbf{0.8995}} & 0.8981       \\ \cline{3-10} 
    & \multicolumn{1}{c|}{}       & \multicolumn{1}{c|}{T1ce}    & 3  & 0.8816& 0.8271& 0.8708       & 0.8825& {\color[HTML]{6434FC} \textbf{0.8856}} & 0.8888       \\ \cline{3-10} 
    & \multicolumn{1}{c|}{\multirow{-4}{*}{BraTS2021}}          & \multicolumn{1}{c|}{FLAIR}   & 4  & 0.9165& 0.8956& 0.9078       & 0.9146& {\color[HTML]{6434FC} \textbf{0.9220}} & 0.9229       \\ \cline{2-10} 
    & \multicolumn{1}{c|}{ATLAS2.0}           & \multicolumn{1}{c|}{T1}      & 5  & 0.5685& 0.3354& 0.5561       & 0.5719& {\color[HTML]{6434FC} \textbf{0.6203}} & 0.5645       \\ \cline{2-10} 
    & \multicolumn{1}{c|}{}       & \multicolumn{1}{c|}{FLAIR}   & 6  & 0.6726& 0.5124& 0.6681       & 0.6723& {\color[HTML]{6434FC} \textbf{0.6909}} & 0.7341       \\ \cline{3-10} 
    & \multicolumn{1}{c|}{\multirow{-2}{*}{WMH2017}}& \multicolumn{1}{c|}{T1}      & 7  & 0.5387& 0.3684& 0.5528       & 0.5511& {\color[HTML]{6434FC} \textbf{0.5773}} & 0.5677       \\ \cline{2-10} 
    & \multicolumn{1}{c|}{}       & \multicolumn{1}{c|}{T1}      & 8  & 0.5638& 0.4354& 0.5932       & 0.5626& {\color[HTML]{6434FC} \textbf{0.5937}} & 0.6028       \\ \cline{3-10} 
    & \multicolumn{1}{c|}{\multirow{-2}{*}{OASIS}}  & \multicolumn{1}{c|}{FLAIR}   & 9  & 0.7241& 0.5753& 0.7431       & 0.7261& {\color[HTML]{6434FC} \textbf{0.7737}} & 0.7970       \\ \cline{2-10} 
    & \multicolumn{1}{c|}{ISLES2022}          & \multicolumn{1}{c|}{DWI}     & 10 & 0.7415& 0.6510& 0.7510       & 0.7368& {\color[HTML]{6434FC} \textbf{0.7723}} & 0.7824       \\ \cline{2-10} 
    & \multicolumn{1}{c|}{}       & \multicolumn{1}{c|}{T1}      & 11 & 0.4664& 0.2426& {\color[HTML]{6434FC} \textbf{0.5408}}     & 0.4630& 0.5044   & 0.5512       \\ \cline{3-10} 
    & \multicolumn{1}{c|}{}       & \multicolumn{1}{c|}{T2}      & 12 & 0.4219& 0.2501& 0.4804       & 0.4781& {\color[HTML]{6434FC} \textbf{0.5384}} & 0.4933       \\ \cline{3-10} 
    & \multicolumn{1}{c|}{}       & \multicolumn{1}{c|}{T1ce}    & 13 & 0.4375& 0.1874& {\color[HTML]{6434FC} \textbf{0.5445}}     & 0.4646& 0.5099   & 0.5586       \\ \cline{3-10} 
\multirow{-14}{*}{\begin{tabular}[c]{@{}c@{}}Seen \\ datasets\end{tabular}}  & \multicolumn{1}{c|}{\multirow{-4}{*}{MSSEG}} & \multicolumn{1}{c|}{FLAIR}   & 14 & 0.6619& 0.4176& 0.6584       & 0.6600& {\color[HTML]{6434FC} \textbf{0.6979}} & 0.7199       \\ \hline
    & \multicolumn{1}{c|}{Tumour1} & \multicolumn{1}{c|}{FLAIR}   & 16 & 0.8358& 0.8339& -& 0.8446& {\color[HTML]{6434FC} \textbf{0.8545}} & 0.8518       \\ \cline{2-10} 
    & \multicolumn{1}{c|}{Tumour2}  & \multicolumn{1}{c|}{T2}      & 15 & 0.7959& 0.8146& -& 0.7968& {\color[HTML]{6434FC} \textbf{0.8293}} & 0.8157       \\ \cline{2-10} 
\multirow{-3}{*}{\begin{tabular}[c]{@{}c@{}}Unseen \\ datasets\end{tabular}} & \multicolumn{1}{c|}{WMHsMix} & \multicolumn{1}{c|}{FLAIR}   & 17 & {\color[HTML]{6434FC} \textbf{0.7032}}       & 0.5479& -& 0.6968& 0.7015   & 0.7508       \\ \hline
    & \multicolumn{2}{c|}{Tumour}   & \begin{tabular}[c]{@{}c@{}}1,2,3,4\\ 15,16\end{tabular}     & 0.8710& 0.8370& -& 0.8732& {\color[HTML]{6434FC} \textbf{0.8825}} & 0.8838       \\ \cline{2-10} 
    & \multicolumn{2}{c|}{WMH}    & \begin{tabular}[c]{@{}c@{}}6,7,8,9,10,\\ 11,12,13,\\ 14,17\end{tabular} & 0.5767& 0.3930& -& 0.5861& {\color[HTML]{6434FC} \textbf{0.6208}} & 0.6417       \\ \cline{2-10} 
\multirow{-5}{*}{\begin{tabular}[c]{@{}c@{}}Lesion\\  level\end{tabular}}   & \multicolumn{2}{c|}{Stroke}  & 5,10  & 0.6550& 0.4932& 0.6535       & 0.6544& {\color[HTML]{6434FC} \textbf{0.6963}} & 0.6734       \\ \hline
    & \multicolumn{2}{c|}{T1}      & 1,5,7,8,11        & 0.6082& 0.4511& 0.6273       & 0.6081& {\color[HTML]{6434FC} \textbf{0.6424}} & 0.6440       \\ \cline{2-10} 
    & \multicolumn{2}{c|}{T2}      & 2,12,15           & 0.7015& 0.6416& -& 0.7214& {\color[HTML]{6434FC} \textbf{0.7557}} & 0.7357       \\ \cline{2-10} 
    & \multicolumn{2}{c|}{T1ce}    & 3,13  & 0.6595& 0.5072& {\color[HTML]{6434FC} \textbf{0.7076}}     & 0.6736& 0.6977   & 0.7237       \\ \cline{2-10} 
\multirow{-4}{*}{\begin{tabular}[c]{@{}c@{}}Modality \\ Level\end{tabular}} & \multicolumn{2}{c|}{FLAIR}   & \begin{tabular}[c]{@{}c@{}}4,6,9,\\ 14,16,17\end{tabular}   & 0.7523& 0.6305& -& 0.7524& {\color[HTML]{6434FC} \textbf{0.7734}} & 0.7961       \\ \hline
\end{tabular}
}
\end{table}
\end{center}

\begin{table}[h]
\centering
\caption{An overview of the public and in-house datasets used for brain lesion segmentation.}
\label{tab:dataset}
\begin{threeparttable}
\resizebox{\textwidth}{!}{
\begin{tabular}{|c|cccccc|ccc|}
\hline
\multirow{2}{*}{Datasets}  
& \multicolumn{6}{c|}{Public datasets}
& \multicolumn{3}{c|}{In-house datasets}  \\ \cline{2-10} 
& \multicolumn{1}{c|}{\href{https://fcon_1000.projects.nitrc.org/indi/retro/atlas.html}{ATLAS2.0}}
& \multicolumn{1}{c|}{\href{https://www.isles-challenge.org/}{ISLES2022}}   
& \multicolumn{1}{c|}{\href{http://www.brainTumoursegmentation.org/}{BraTS2021}} 
& \multicolumn{1}{c|}{\href{https://wmh.isi.uu.nl/\#_Toc122355653}{WMH2017}}   
& \multicolumn{1}{c|}{\href{https://portal.fli-iam.irisa.fr/msseg-challenge/}{MSSEG}}  
& {\href{https://www.oasis-brains.org/}{OASIS}}   
& \multicolumn{1}{c|}{Tumour1}  
& \multicolumn{1}{c|}{Tumour2}  
& WMHsMix\\ \hline
Lesion type& \multicolumn{1}{c|}{\begin{tabular}[c]{@{}c@{}}Chronic \\ stroke\end{tabular}} & \multicolumn{1}{c|}{\begin{tabular}[c]{@{}c@{}}Ischemic \\ stroke\end{tabular}} & \multicolumn{1}{c|}{\begin{tabular}[c]{@{}c@{}}Glioma \\ tumour\end{tabular}} & \multicolumn{1}{c|}{WMH}  
& \multicolumn{1}{c|}{MS}  
& \begin{tabular}[c]{@{}c@{}}Normal \\ healthy\end{tabular} & \multicolumn{1}{c|}{\begin{tabular}[c]{@{}c@{}}Glioma\\ tumour\end{tabular}} & \multicolumn{1}{c|}{\begin{tabular}[c]{@{}c@{}}Glioma\\ tumour\end{tabular}} & \begin{tabular}[c]{@{}c@{}} WMH Mixture\footnote[3]{}: \\
PD,
NMOSD, \\
MS,
SVD,
AD
\end{tabular}\\ \hline
Modality  
& \multicolumn{1}{c|}{T1}  
& \multicolumn{1}{c|}{DWI}
& \multicolumn{1}{c|}{\begin{tabular}[c]{@{}c@{}}T1, T1ce\footnote[1]{},\\T2, FLAIR\end{tabular}} & \multicolumn{1}{c|}{\begin{tabular}[c]{@{}c@{}}T1,\\ FLAIR\end{tabular}} & \multicolumn{1}{c|}{\begin{tabular}[c]{@{}c@{}}T1, T1ce,\\T2, FLAIR\end{tabular}} 
& \begin{tabular}[c]{@{}c@{}}T1,\\ FLAIR\footnote[2]{}\end{tabular}  
& \multicolumn{1}{c|}{T2} 
& \multicolumn{1}{c|}{FLAIR}   
& FLAIR   \\ \hline
\begin{tabular}[c]{@{}c@{}}\textbf{\#}Training\\images\end{tabular} & \multicolumn{1}{c|}{545}  
& \multicolumn{1}{c|}{150}
& \multicolumn{1}{c|}{4168}
& \multicolumn{1}{c|}{60}   
& \multicolumn{1}{c|}{36}  
& 200  
& \multicolumn{1}{c|}{29\footnote[4]{}}
& \multicolumn{1}{c|}{27\footnote[4]{}}
& \multicolumn{1}{c|}{20\footnote[4]{}}\\ \hline
\begin{tabular}[c]{@{}c@{}}\textbf{\#}Test\\images\end{tabular} 
& \multicolumn{1}{c|}{110} 
& \multicolumn{1}{c|}{100}
& \multicolumn{1}{c|}{836} 
& \multicolumn{1}{c|}{60}   
& \multicolumn{1}{c|}{24}  
& 100  
& \multicolumn{1}{c|}{40} 
& \multicolumn{1}{c|}{40} 
& 40 \\ \hline
\end{tabular}
}
 \begin{tablenotes}
\footnotesize
\item[1] Annotations for tumour core.
\item[2] In-house expert annotations based on FLAIR.
\item[3] Parkinson's Disease (PD), Neuromyelitis Optica Spectrum Disorder (NMOSD),\\Multile Sclerosis (MS), Small Vessel Disease (SVD), Alzheimer's Disease (AD).
\item[4] Not used for foundation model training, but used for training task-specific models\\regarded as upperbound for generalisation onto unseen datasets.
\end{tablenotes}  
\end{threeparttable}   
\end{table}

\begin{figure}[h]
\centering
\includegraphics[width=\textwidth]{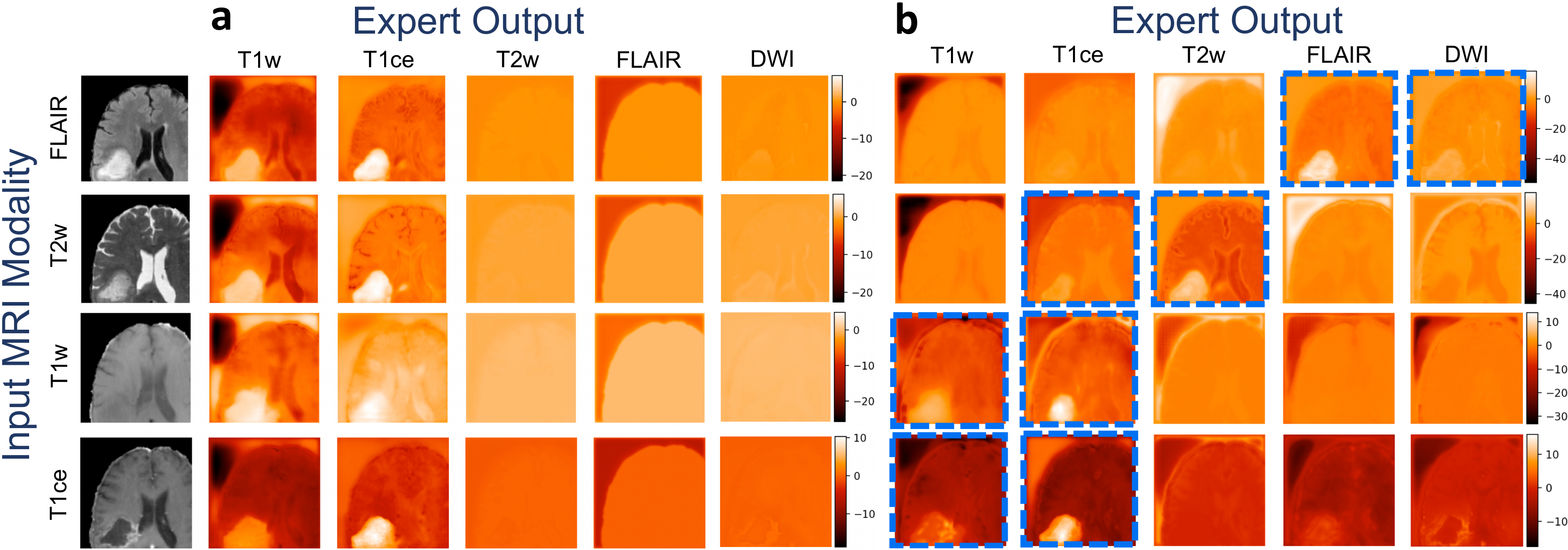}
\caption{The output of each modality expert weighted by the gating mechanism achieved a) without or b) with the proposed anti-degeneration curriculum learning strategy. 
In a), some experts lose their acquired modality knowledge and the activation of the experts becomes imbalanced, whereas in b),  expert degeneration is avoided and each expert is activated by its specialised modality or related modalities that correlate with the specialised modality (see blue dashed boxes).
}
\label{fig:second_pdf}
\end{figure}